\documentclass[aps,prl,twocolumn,showpacs,superscriptaddress]{revtex4}
\usepackage{graphicx}
\usepackage{epsfig}

\def\Ket#1{\left|#1\right>}
{\catcode`\|=\active
  \gdef\Braket#1{\left<\mathcode`\|"8000\let|\bravert {#1}\right>}}
\def\bravert{\egroup\,\vrule\,\bgroup}

\begin{document}

\title{Four-Wave mixing in degenerate Fermi gases: Beyond the undepleted pump
approximation}
\author{T. Miyakawa}
\affiliation{Optical Sciences Center, The University of Arizona,
Tucson, AZ 85721}
\affiliation{Department of Physics, Tokyo Metropolitan University,
1-1 Minami-Ohsawa, Hachioji, Tokyo 192-0397, Japan}
\author{H. Christ}
\author{C. P. Search}
\author{P. Meystre}
\affiliation{Optical Sciences Center, The University of Arizona,
Tucson, AZ 85721}
\date{\today}

\begin{abstract}
We analyze the full nonlinear dynamics of the four-wave mixing
between an incident beam of fermions and a fermionic density
grating. We find that when the number of atoms in the beam is
comparable to the number of atoms forming the grating, the
dephasing of that grating, which normally leads to a decay of its
amplitude, is suppressed. Instead, the density grating and the
beam density exhibit large nonlinear coupled amplitude
oscillations. In this case four-wave mixing can persist for much
longer times compared to the case of negligible back-action. We
also evaluate the efficiency of the four-wave mixing and show that
it can be enhanced by producing an initial density grating with an
amplitude that is less than the maximum value. These results
indicate that efficient four-wave mixing in fermionic alkali gases
should be experimentally observable.
\end{abstract}

 \maketitle

\section{Introduction}
In recent years the field of nonlinear atom optics \cite{Lenz} has
become firmly established both theoretically and experimentally
with the demonstration of four-wave mixing in Bose-Einstein
condensates \cite{Deng99}, coherent matter wave amplification
\cite{amplification}, the creation of dark \cite{Burg99} and
bright \cite{Stre02} atomic solitons, the generation of correlated
atomic pairs and squeezing \cite{Voge02,Voge02-2}, and the
creation of a molecular condensate -- the analog of
second-harmonic generation in optics \cite{Wyna00,Mcke02}. So far,
this work has been largely limited to bosonic atoms and has been
based to a great extent on analogies with nonlinear optics:
Ultracold bosons interact primarily via $s$-wave collisions and
typically occupy only a few quantum states, so that their
mean-field description in terms of a Gross-Pitaevskii equation
includes a nonlinearity of the same form as a third-order optical
Kerr nonlinearity, and thus many of the results from nonlinear
optics can be directly applied to condensates. In particular,
Bose-Einstein condensates are often viewed as a matter-wave
equivalent of the optical laser-- the so called "atom laser".

The last few years have also seen substantial progress in the
cooling and trapping of atomic Fermi gases to the quantum
degenerate regime, with temperatures reaching as low as $T <
0.2T_F$ where $T_F$ is the Fermi temperature
\cite{Lithium1,Dema99,Hadz02,Roat02,OHara02}. It is therefore only
a matter of time before experiments start to probe the dynamics of
these gases far from equilibrium and perform fermionic nonlinear
atom optics experiments. However, the extension of the ideas
developed in nonlinear optics to Fermi systems is by no means
obvious or straightforward because of the different quantum
statistics obeyed by fermions and bosons. Whereas for bosons the
analysis can typically be limited to a few macroscopically
occupied modes, the Pauli exclusion principle prohibits the number
of modes from ever being less than the number of particles for
fermions. This makes both analytical and numerical studies of the
dynamics of Fermi systems rather challenging, except for very
limited situations (e.g. $T=0$ and weak perturbations).
Consequently, the idea of doing nonlinear atom optics with
fermions is still new and largely unexplored.

It is only recently that it has been convincingly argued that
four-wave mixing is possible with fermions \cite{Kett01,Moor01}.
These works showed how four-wave mixing can be interpreted in
terms of incident particles undergoing Bragg scattering off of a
fermionic density grating, but were limited to an incident beam
consisting of a single test particle. They also neglected the
dynamics of the density grating. Related work has shown that
four-wave mixing between a degenerate fermion beam and a bosonic
density grating is also possible \cite{Vill01}. While these
results are promising, a complete analysis of purely fermionic
four-wave mixing is still lacking.

In a previous paper \cite{4wm1}, we studied fermionic four-wave
mixing using the matter-wave analog of the undepleted pump
approximation in optics, i.e. we neglected the back-action of the
diffracted atoms on the density grating. The density grating was
formed by atoms in a single hyperfine state prepared in a coherent
superposition of the momentum states $k-q/2$ and $k+q/2$ for all
$k$ such that $|k|<k_{F,g}$, where $k_{F,g}$ is the Fermi wave
number of the grating atoms and $q$ the wave number of the density
modulation. The third incident wave consisted of a degenerate
Fermi gas in a different hyperfine state so that the atoms in the
beam and grating could interact via $s$-wave collisions. These
atoms could then be diffracted by the density grating to produce a
fourth, scattered wave. Because of the range of energy states
occupied by its constituent atoms, the amplitude of the grating
would decay away due to the dephasing of the atoms. This led to a
finite time during which four-wave mixing could occur and placed
serious constraints on the properties of the fermions involved.

In this paper, we extend the results of Ref. \cite{4wm1} by taking
into account the back-action of the incident beam on the fermionic
density grating. We find that when the number of atoms in the
incident beam becomes a sizable fraction of the number of atoms
forming the density grating, it no longer decays away due to the
dephasing. Instead, it exhibits large nonlinear amplitude
oscillations that are coupled to the Bragg oscillations of the
beam. This leads to the efficient generation of the fourth
scattered wave, even for times much longer than the grating
dephasing time. We interpret this behavior in terms of a Bloch vector
picture. We also evaluate the efficiency for the generation of the
scattered wave and show that it should be possible to produce a
macroscopic fourth wave. Interestingly enough, we find that the
efficiency of the four-wave mixing can be improved significantly
by initially preparing a density grating with an amplitude that is
less than maximal.

The organization of this paper is as follows: In section II, we
briefly review the model developed in \cite{4wm1}. Section III
presents numerical results that show how the dephasing of the
grating can be eliminated when the back-action becomes
significant. The efficiency of the four-wave mixing is also
calculated and it is shown that a macroscopic fourth wave can be
readily produced. Finally, section IV is a summary and conclusion.
The appendix presents an analytically soluble model in terms of
coupled Bloch vectors that yields a qualitative interpretation of
our results.

\section{model}
\begin{figure}
\begin{center}
\includegraphics*[width=8cm,height=5cm]{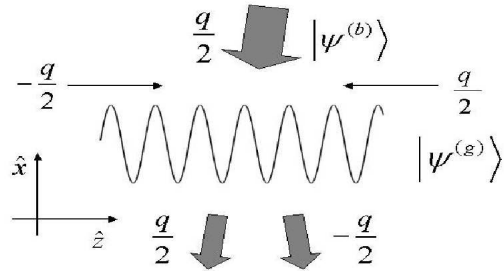}
\vspace{.3 cm}
 \caption{Model of the scattering. The spin-down polarized beam is
 scattered ($s$-wave) by a spin-up polarized grating.}
  \label{model}
\end{center}
\end{figure}

In this section we briefly review the one-dimensional model
developed in \cite{4wm1} to study four-wave mixing between
fermions at zero temperature. We consider two spin-polarized
species of fermions-- a density grating composed of spin-up
polarized atoms and an incident beam of spin-down polarized atoms,
see Fig.~\ref{model}. Atoms in different spin states interact via
elastic $s$-wave scattering. The cross-section for $p$-wave
scattering at $T=0$ is orders of magnitude smaller than that of
$s$-wave scattering, thus $p$-wave scattering is neglected in our
model \cite{Dema98}.

For clarity and simplicity we restrict ourselves to one spatial
dimension. It is usually possible to reduce the problem of
four-wave mixing to a two-dimensional planar geometry, with one
dimension parallel to the density grating. Since the momentum
transfer to the beam atoms is parallel to the density modulation
of the grating, the essential dynamics of the wave mixing process
occurs only in that $\hat{z}$-direction. We assume that the
grating is confined between $x=0$ to $x=d$ along the perpendicular
$\hat{x}$ axis. If the average momentum, $\hbar k_{\perp}$, of the atomic
beam along that direction is sufficiently large that $k_{\perp}\gg
1/d$, then the requirements of energy and momentum conservation
insure that $k_\perp$ remains practically unchanged by the
scattering. In this case, wave mixing in two or three dimensions
is in essence the same as in the one-dimensional model that we
consider. (The transverse momenta of the beam atoms can at most
change sign due to reflection off the grating at $x=0$
\cite{Vill01}, but the reflected wave can be minimized by using a
smooth grating density profile in the $\hat{x}$-direction.)

The second quantized Hamiltonian describing the situation at hand
is
\begin{eqnarray}\label{BasicHamiltonian}
  H&=&\sum_k \hbar \omega_{k}\left( {a}^{\dag}_{k \uparrow}a_{k
  \uparrow} + {a}^{\dag}_{k \downarrow}a_{k
  \downarrow} \right) \nonumber\\
  &+&\hbar U_0 \sum_{k_1,k_2,q}\left(
  a^{\dag}_{k_1+q \uparrow}a^{\dag}_{k_2-q \downarrow} a_{k_2
  \downarrow}a_{k_1 \uparrow} \right),
\end{eqnarray}
with $\omega_k=\hbar k^2/2m$ and $U_0=4\pi \hbar a/(Ld^2m)$. Here,
$a$ is the scattering length, $d^2$ is the transverse
cross-sectional area, $m$ the mass of one atom and $L$ the
quantization length. The operators  $a_{k,\sigma}$ and
$a_{k,\sigma}^{\dagger}$ are annihilation and creation operators
for atoms with momentum $\hbar k$ and spin
$\sigma={\uparrow,\downarrow}$, satisfying standard fermionic
anticommutation relations.

The dynamics of the system is conveniently described in terms of
the particle-hole operators
\begin{eqnarray}\label{particleholeoperators}
  \rho_{k,k'}^{(b)}&=&a^{\dag}_{k+k' \downarrow}a_{k \downarrow},
  \nonumber\\
  \rho_{k,k'}^{(g)}&=&a^{\dag}_{k+k' \uparrow}a_{k \uparrow},
\end{eqnarray}
that create a particle in state $k+k'$ and a hole in $k$ for the
beam and grating, respectively. They obey the Heisenberg equations
of motion
\begin{eqnarray}\label{particleholefullequations}
&&i \frac{d}{dt}\rho_{k,k'}^{(b,g)}=
\omega_{k,k'}\rho_{k,k'}^{(b,g)}\nonumber\\
&+&U_0 \sum_{k_1,k_2}\left(
\rho_{k+k_1,k'-k_1}^{(b,g)}-\rho_{k,k'-k_1}^{(b,g)}
\right)\rho_{k_2,k_1}^{(g,b)},
\end{eqnarray}
with $\omega_{k,k'}=\omega_k - \omega_{k+k'}$.

We consider a beam of $N_b$ atoms propagating along the ${\hat
x}$-direction, but with a small momentum component around a
central value $\overline{q}$ parallel to the grating, which is
taken to be in a superposition of the states $k-q/2$ and $k+q/2$
for $-k_{F,g}\leq k\leq k_{F,g}$. Here, $k_{F,g}=\pi N_g/L$ is the
Fermi momentum (in one dimension) and $N_g$ is the number of atoms
in the grating. Such a grating can be created from a stationary
homogeneous gas by a two-step process where first a two-photon Bragg $\pi$-pulse
imparts a momentum $-\hbar q/2$ to each atom, followed by a second
Bragg pulse with momentum $\hbar q$ that creates the desired
superposition (see e.g. Ref. \cite{Torii}). (Note that the
bandwidth of the pulses should be greater than
$\omega_{k_{F,g}}$.) Thus the initial state of the beam-grating
system is
\begin{equation}
\Ket{\Psi(0)}=\Ket{\Psi^{(b)}}\Ket{\Psi^{(g)}},
\end{equation}
where
\begin{eqnarray}
\label{initialconditions}
 \Ket{\Psi^{(g)}}&=&\prod_{|k|\leq
 k_{F,g}}\frac{1}{\sqrt{2}} \left( a^{\dag}_{k-q/2 \uparrow}+
 a^{\dag}_{k+q/2 \uparrow}\right) \Ket{0},
 \nonumber \\
 \Ket{\Psi^{(b)}}&=& \prod_{|k| \leq k_{F,b}}
 a^{\dag}_{k+\overline{q} \downarrow}\Ket{0}.
\end{eqnarray}
$k_{F,b}=\pi N_b/L$ being the Fermi momentum of the beam.

The initial expectation values of the particle-hole operators
$\rho_{k,k'}^{(b)}$ and $\rho_{k,k'}^{(g)}$ are readily found to
be
\begin{eqnarray}
&&\Braket{\Psi^{(b)} |\rho_{k,k'}^{(b)}(0) |
\Psi^{(b)}}=\delta_{k',0}\Theta(k_{F,b}-|k-\overline{q}|),
\nonumber\\
&&\Braket{\Psi^{(g)} |\rho_{k,k'}^{(g)}(0)
|\Psi^{(g)}}=\frac{1}{2}\left( \delta_{k',0} + \delta_{k',q}
\right)\Theta(k_{F,g}-|k+\frac{q}{2}|)\nonumber\\
&&+\frac{1}{2}\left( \delta_{k',0} + \delta_{k',-q}
\right)\Theta(k_{F,g}-|k-\frac{q}{2}|). \label{initial_state}
\end{eqnarray}
We note that the spatial density of the Fermi gas is given by
\begin{equation}
\rho^{(b,g)}(x)=\frac{1}{L}\sum_{k'}\left(\sum_{k}\langle
\rho^{(b,g)}_{k,k'}\rangle \right)e^{-ik'x},
\end{equation}
from which it is clear that the density for the initial state of
the spin-up polarized grating atoms exhibits as expected a spatial
modulation with the grating wave number $q$ and an amplitude
$N_g/L$. In contrast, the spin-down polarized incident beam has a
uniform density.

It is convenient to introduce the slowly varying particle-hole
operators
$\rho_{k,k'}^{(b,g)}(t)=\hat{\rho}_{k,k'}^{(b,g)}(t)e^{-i\omega_{k,k'}t}$
whose Heisenberg equation of motion simply become
\begin{widetext}
\begin{equation}
\frac{d}{dt}\hat{\rho}_{k,k'}^{(b,g)} = -i U_0 \sum_{k_1,k_2}
\left( \hat{\rho}_{k+k_1,k'-k_1}^{(b,g)}
e^{i(\omega_{k,k_1}-\omega_{k_2,k_1})t}-
\hat{\rho}_{k,k'-k_1}^{(b,g)}
e^{-i(\omega_{k+k',-k_1}+\omega_{k_2,k_1})t}
\right)\hat{\rho}_{k_2,k_1}^{(g,b)}(t),\label{integrated}
\end{equation}
In the following we factorize the expectation values of these
operators for different spin states,
\begin{equation}
\langle \hat{\rho}_{k,q}^{(g)}\hat{\rho}_{k',k''}^{(b)}\rangle
\approx \langle \hat{\rho}_{k,q}^{(g)}\rangle \langle
\hat{\rho}_{k',k''}^{(b)}\rangle. \label{factorize}
\end{equation}
This ansatz results in the truncation of the BBGKY-type hierarchy
of equations of motion for the higher-order moments of the
particle-hole operators after the first moments \cite{Huang},
yielding the closed set of $c$-number equations
\begin{equation}
\frac{d}{dt}\bar{\rho}_{k,k'}^{(b,g)} = -i U_0 \sum_{k_1,k_2}
\left( \bar{\rho}_{k+k_1,k'-k_1}^{(b,g)}
e^{i(\omega_{k,k_1}-\omega_{k_2,k_1})t}-
\bar{\rho}_{k,k'-k_1}^{(b,g)}
e^{-i(\omega_{k+k',-k_1}+\omega_{k_2,k_1})t}
\right)\bar{\rho}_{k_2,k_1}^{(g,b)}(t), \label{integrated_c}
\end{equation}
\end{widetext}
where
$\bar{\rho}_{k,k'}^{(g)}\equiv\langle\hat{\rho}_{k,k'}^{(g)}\rangle$
and
$\bar{\rho}_{k,k'}^{(b)}\equiv\langle\hat{\rho}_{k,k'}^{(b)}\rangle$.
These equations can be readily integrated numerically.

Implicit in Eqs. (\ref{integrated_c}) and (\ref{initial_state})
are the three fundamental time scales that govern the dynamics of
the interaction between the beam and the grating. The first time
scale is associated with the kinetic energy $\omega_q$ gained by
an atom scattered by the grating. The other two time scales are
associated with the density grating itself. If, for the moment, we
revert to the undepleted pump approximation used in Ref.
\cite{4wm1}, namely
$\bar{\rho}_{k,q}^{(g)}(t)=\bar{\rho}_{k,q}^{(g)}(0)$, then we
obtain the linear equations of motion for the beam,
\begin{eqnarray}\label{lin}
\frac{d}{dt} \bar\rho_{k,k'}^{(b)}&=& -ig(t)\left(
\bar\rho_{k+q,k'-q}^{(b)}e^{i\omega_{k,q}t}-
\bar\rho_{k,k'-q}^{(b)}e^{-i\omega_{k+k',-q}t} \right. \nonumber\\
&-&\left.\bar\rho_{k-q,k'+q}^{(b)}e^{i\omega_{k,-q}t}+
\bar\rho_{k,k'+q}^{(b)}e^{-i\omega_{k+k',q}t} \right).
\end{eqnarray}
Here
\begin{eqnarray}
\label{timedepcoupling}
g(t)&=&\frac{U_0}{2}\sum_{k_2}\Theta(k_{F,g}-|k_2+q/2|)e^{-i\omega_{k_2,q}t} \nonumber\\
&=&\frac{U_0N_g}{2}{\rm sinc}\left(\frac{\hbar
qk_{F,g}t}{m}\right) ,\nonumber
\end{eqnarray}
where the second equality was obtained in the continuum limit,
$\sum_k \rightarrow L/2 \pi \int dk$. The coupling constant $g(t)$
is proportional to the amplitude of the density grating, which
decays to zero in a time of the order of
\[
\tau_D=m/\hbar qk_{F,g}.
\]
This decay is caused by the spread in energies of the fermions
that comprise the grating,
$\hbar\Delta\omega=\hbar|\omega_{k_{F,g}+q/2}-\omega_{k_{F,g}-q/2}|$,
so that the different energy states in the grating get out of
phase with each other over time. This leads to an effective
grating lifetime of the order of $\tau_D=1/\Delta\omega$, which
corresponds to the relative phases of the different states being
of order unity.

The third time scale is given by the characteristic time
$$\tau_B=1/U_0N_g$$ for atoms to scatter off of the grating in
the limit of a static grating ($\tau_D\rightarrow\infty$). For an
incident particle with momentum $\pm q/2$, $\tau_B^{-1}$ is the
frequency of oscillations between the degenerate states $q/2$ and
$-q/2$. In the Bragg regime that we consider in this paper, both
the kinetic energy and the momentum of the atoms are conserved by
the scattering process. This corresponds to the condition,
\[
\omega_q\tau_B\gg 1.
\]
The population that develops in states centered around $\pm
3q/2,\pm 5q/2,\ldots$ is then at most on the order of
$(\tau_B\omega_q)^{-2}$. If we choose the average momentum of the
beam as $\bar{q}=q/2$, with $\bar{q}>k_{F,b}$, only states
centered around $-q/2$ are phase-matched to the states in the
incident beam, so that we can restrict ourselves to those $k$
states within a width $2k_{F,b}$ of $\pm q/2$. (This is in contrast
to the Raman-Nath regime, where the change in the kinetic energy
of an atom as result of scattering off of the grating is
negligible compared to the interaction energy with the grating
\cite{Rojo99}.)

In the undepleted pump approximation it is important that the
scattering time be less than the lifetime of the grating,
\begin{equation}
\tau_B<\tau_D ,\label{nodephase}
\end{equation}
otherwise there is no significant scattering beam generated.
However, as we show in the next section, four-wave mixing can
continue for times much longer than $\tau_D$ when the effects of
the back-action on the grating become significant.

\section{Numerical Results}
In this section we study the phenomena in fermionic four-wave
mixing resulting from the back-action of the atomic beam on the
grating. We proceed by numerically solving Eqs.
(\ref{integrated_c}), subject to the initial conditions
(\ref{initial_state}) with $\bar{q}=q/2 >k_{F,b}$, using a
fourth-order Runge-Kutta algorithm with fixed time steps chosen to
produce a relative error smaller than $1\%$.
\begin{figure}
\begin{center}
\includegraphics[width=8cm,height=11cm]{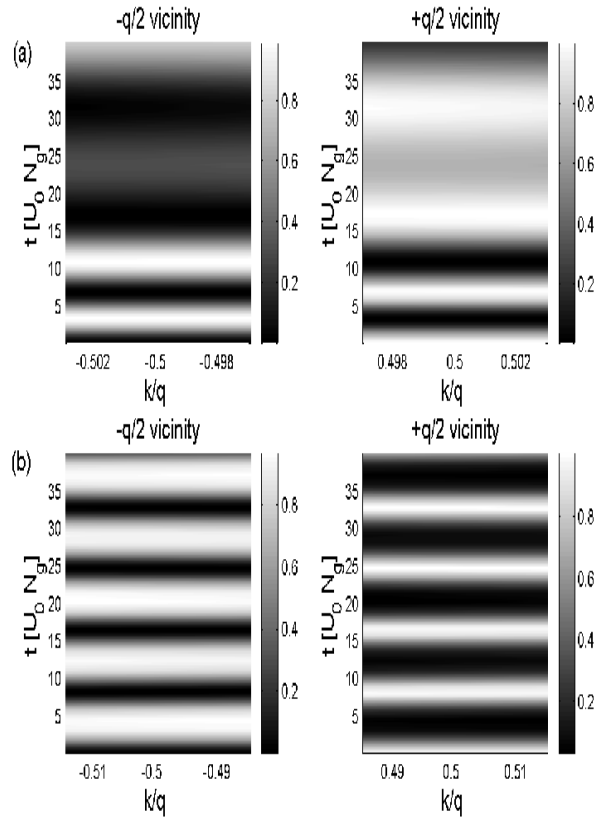}
\vspace{.3 cm}
 \caption{Populations $n^{(b)}_k$ as a function of time (in units of
 $\tau_B=1/U_0 N_g$) for (a) $N_b/N_g=0.1$ and (b) $N_b/N_g=0.5$. In both
 cases $\tau_D/\tau_B=7.5$ and $\omega_q/(U_0 N_g)$=2.}
  \label{fig2}
\end{center}
\end{figure}

The back-action of the beam on the density grating is negligible
for all times when the number of beam atoms is much less than the
number of atoms in the grating, $N_g\gg N_b$. The reason is that
for every atom in the beam that suffers a momentum transfer of
$\pm \hbar q$, one atom of the grating sees its wave function
``reduced'' from the superposition state
$(|k-q/2\rangle+|k+q/2\rangle)/\sqrt{2}$ to the state $|k\mp q/2
\rangle$. Since only those atoms in a superposition of $k-q/2$ and
$k+q/2$ (for all possible $k$) contribute to the density
modulation, the fractional reduction in the amplitude of the
grating will be at most $N_b/N_g\ll 1$ and the undepleted pump
approximation is valid for all times.  This approximation also
holds for arbitrary $N_b$ for times short enough that the number
of atoms scattered by the grating is much less than $N_g$, that
is, for times $t\ll \tau_B$.
\begin{figure}
\begin{center}
\includegraphics[width=8cm,height=7cm]{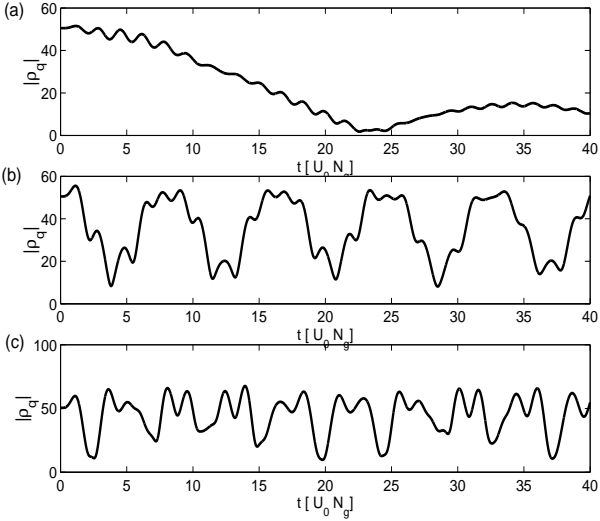}
\vspace{.3 cm}
 \caption{The densities $|\rho_q|$ as a function of time scaled by $U_0 N_g$ for
 the ratios $N_b/N_g=0.1,0.5$ and $1.0$ in the subplot (a), (b) and (c), respectively. }
  \label{fig3}
\end{center}
\end{figure}

Fermionic four-wave mixing is conveniently characterized in terms
of the occupation numbers of the momentum states of the beam and
grating atoms
\begin{equation}
n^{(b,g)}_k(t)=\bar{\rho}^{(b,g)}_{k,0}(t),
\end{equation}
and the amplitude of the density modulation of the grating,
\begin{equation}
\rho_q(t)=\sum_k\bar{\rho}_{k,q}^{(g)}(t)e^{-i\omega_{k,q}t}.
\end{equation}

Fig.~\ref{fig2}a plots $n^{(b)}_k$ as a function of time for the
case of a ``large'' grating, $N_b/N_g=0.1$ and
$\tau_D/\tau_B=7.5$, and Fig.~\ref{fig3}a shows the evolution of
the density modulation of the grating $|\rho_q(t)|$ for these same
parameters. The atoms in the beam undergo Bragg oscillations
between the $k$ states centered around $-q/2$ and $+q/2$ up until
a time of the order of $\tau_D$, when the amplitude of the
oscillations dies out. It is clear from the behavior of
$|\rho_q(t)|$ that the disappearance of the Bragg oscillations
corresponds to the decay of the grating amplitude due to
dephasing. In this example, the back-action of the beam on the
grating manifests itself in the small amplitude oscillations
superimposed on the sinc function behavior of $|\rho_q(t)|$. These
oscillations reflect the coupling between individual modes of the
beam and of the grating, which all occur at roughly the same
frequency $\tau_B^{-1}$ for $\tau_B^{-1} \gg |kq/m|$, as further
discussed later on.

\begin{figure}
\begin{center}
\includegraphics[width=8cm,height=11cm]{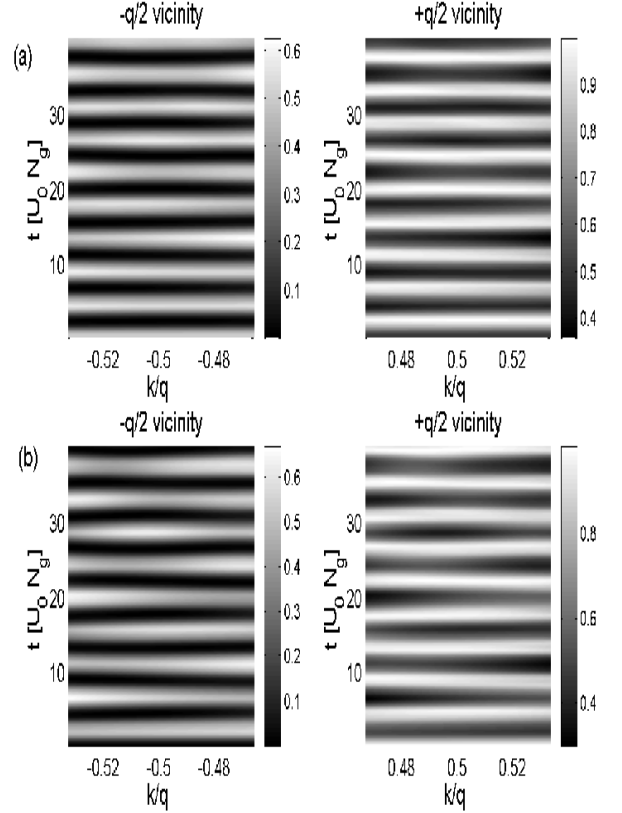}
\vspace{.3 cm}
 \caption{Atomic populations as a function of time (in units of
 $\tau_B=1/U_0 N_g$) for the case of equal particle numbers
 $N_b/N_g=1.0$. 
 The upper plots (a) show the density of the grating, $n_k^{(g)}$,
 and the lower plots (b) show the beam evolution, $n_k^{(b)}$.
 }
  \label{fig4}
\end{center}
\end{figure}

Figs.~\ref{fig2}b and \ref{fig4}b are the same as Fig.~\ref{fig2}a
except that ``smaller'' gratings, with $N_b/N_g=0.5$ and
$N_b/N_g=1.0$, respectively, have been used. For these parameters
there is a dramatic change in the behavior of $|\rho_q(t)|$ as the
back-action becomes more pronounced. Instead of decaying away, the
density grating now undergoes large amplitude oscillations, as
illustrated in Fig. \ref{fig3}. At the same time, the beam atoms
undergo nearly perfect Bragg oscillations that persist for times
much longer than $\tau_D$. Note that the amplitude of the
oscillations is equal to 1 for $N_b/N_g=0.5$, but only to about
$1/2$ for $N_b/N_g=1.0$ (see the scales of the grey-scale plots)
although there is no evidence of any decay in the Bragg
oscillations.

If one ignores the weak population of the states centered around $\pm 3q/2,\pm
5q/2,\ldots$ (an approximation valid in the Bragg regime) these
coupled oscillations involve the oscillation of a beam atom
between the states $|k'+q/2\rangle$ and $|k'-q/2\rangle$, while an
atom in the grating oscillates between $|k-q/2\rangle$ and
$|k+q/2\rangle$, $\pi/2$ out of phase with the beam atom, as
follows from the existence of the constants of motion
\begin{eqnarray}
&&\sum_{|k|<k_{F,g}}n^{(g)}_{k+q/2}+\sum_{|k|<k_{F,b}}n^{(b)}_{k+q/2},
\label{constant1} \\
&&\sum_{|k|<k_{F,g}}n^{(g)}_{k-q/2}+\sum_{|k|<k_{F,b}}n^{(b)}_{k-q/2}.
\label{constant2}
\end{eqnarray}

We can gain a better understanding of fermionic four-wave mixing
by recasting it in terms of the dynamics of an inhomogeneously
broadened ensemble of coupled two-level systems. In this picture,
an individual two-level system consists of the manifold of beam
and grating states $\{|k-q/2\rangle, |k+q/2\rangle \}$ for all
states with $|k|<k_{F,g}$ and $|k|<k_{F,b}$ for the grating and
beam atoms, respectively. We proceed by introducing the associated
Bloch vectors \cite{quantumoptics},
\begin{equation}
{\bf U}^{(g)}_k=U_k^{(g)}{\bf \hat{x}}+V_k^{(g)}{\bf
\hat{y}}+W_k^{(g)}{\bf \hat{z}} \label{Bvector_g},
\end{equation}
with the components
\begin{eqnarray}
U^{(g)}_k(t)&=&\rho_{k-q/2,q}^{(g)}(t)+\rho_{k+q/2,-q}^{(g)}(t), \label{Ug}\\
V^{(g)}_k(t)&=&i\rho_{k-q/2,q}^{(g)}(t)-i\rho_{k+q/2,-q}^{(g)}(t), \label{Vg}\\
W^{(g)}_k(t)&=&n^{(g)}_{k+q/2}-n^{(g)}_{k-q/2} \label{Wg}.
\end{eqnarray}
It is easily seen that the Bloch vectors associated with the atoms
in the incident beam obey the equations of motion
\begin{equation}
\frac{d}{dt}{\bf U}^{(g)}_k(t)={\bf U}^{(g)}_k(t)\times{\bf
R}_k^{(b)}(t) \label{Bloch},
\end{equation}
that is, they precess around the effective field
\begin{equation}
{\bf R}_k^{(b)}(t)=U_0\sum_{k'} U_{k'}^{(b)}{\bf
\hat{x}}+U_0\sum_{k'} V_{k'}^{(b)}{\bf \hat{y}}+\delta\omega_k{\bf
\hat{z}},
\end{equation}
where
\begin{equation}
\delta\omega_k=\frac{\hbar kq}{m}.
\end{equation}
Here, $U_{k'}^{(b)}$ and $V_{k'}^{(b)}$ are the components of the
$k'$-th Bloch vector for the beam, ${\bf U}^{(b)}_{k'}$. They are
defined in the same way as Eqs. (\ref{Bvector_g})-(\ref{Wg}) for
the grating. Clearly, the grating Bloch vector ${\bf U}_k^{(b)}$
also obeys Eq. (\ref{Bloch}), with the interchange
$b\leftrightarrow g$.

The components of the Bloch vector in the equatorial
$(\hat{x}\hat{y})$ plane of the Bloch sphere are a measure of the
coherence between the momentum states $|k-q/2\rangle$ and
$|k+q/2\rangle$ and therefore the sum of the projections of the
grating Bloch vectors onto the equatorial plane is equal to the
amplitude of the density grating.
\begin{figure}
\begin{center}
\includegraphics[width=8cm,height=5cm]{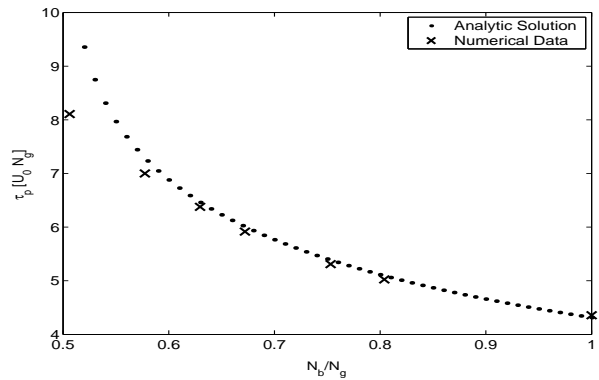}
\vspace{.3 cm}
 \caption{Crosses: numerically obtained period of oscillation $\tau_p$ of the
 atomic densities; Dotted line: analytic result of Eq. (\ref{period}). Time in units of $U_0N_g$. }
  \label{fig5}
\end{center}
\end{figure}

Initially all of the Bloch vectors for the grating point along the
$\hat{x}$ axis, while those of the atomic beam, which doesn't
exhibit any coherence, point toward the north pole ($+\hat{z}$) of
the Bloch sphere. (That they point north rather than south follows
from the fact that the initial momenta of the beam are in an
interval of width $k_{F,b}$ about ${\bar q}=q/2$, see Eq.
(\ref{initial_state}). We have seen that the Bragg regime requires
additionally that ${\bar q} > k_{F,b}$, so that $n_{k-q/2}^{(b)}$
is initially equal to zero and $W_k^{(b)} > 0$.)

In the absence of any coupling to the beam, $U_0 =0$, the version
of Eq. (\ref{Bloch}) for the grating shows that the evolution of
the Bloch vectors is simply a rotation in the equatorial plane at
frequency $\delta \omega_k$. This ``inhomogeneous broadening''
causes the grating to dephase because of the spreading out of the
Bloch vectors in the equatorial plane.

Consider now the other extreme situation where the collisional
coupling between the beam and grating atoms completely dominates
the dynamics, i.e. $U_0 N_b \sim U_0 N_g \sim U_0\sum_k
U^{(g)}_k(t) \gg \delta \omega_k$ or equivalently, $\tau_D \gg \tau_B$. The amplitude of the effective
field that drives the grating, $U_0\sum_k V_k^{(b)}(t)$, is
initially equal to zero since the initial Bloch vectors of the
beam point towards the north pole ($+\hat{z}$) of the Bloch
sphere. However, those vectors start rotating around the
$\hat{x}$-axis under the influence of the instantaneous grating
effective field $U_0\sum_kU_k^{(g)}(t)$, leading in turn to a
projection along the ${\hat y}$-axis. This then allows the Bloch
vector of the grating to start rotating about that axis towards
the north pole of the Bloch sphere at the instantaneous frequency
$U_0\sum_kV_k^{(b)}(t)$. As ${\bf U}_k^{(b)}$ approaches the
${\hat y}$-axis, ${\bf U}_k^{(g)}$ reaches the north pole and its
projection on the equatorial plane vanishes and then changes sign.
This reverses the sense of rotation of the beam Bloch vectors,
which begin to move back toward the pole, at which point their
projections on the ${\hat y}$-axis likewise change sign. This
reverses the direction of rotation of ${\bf U}_k^{(g)}$, which
then passes back through the north pole, reversing the sense of
rotation of ${\bf U}_k^{(b)}$, etc. We see that the beam atoms are
transformed into a grating-like superpositions state, and at the
same time the atoms forming the initial grating take on a state
similar to the initial beam state. Thus, as time passes, the
labels ``beam'' and ``grating'' become more and more
interchangeable. One result of this back-and-forth is that the
oscillation frequency of the populations is twice that of the
coherences. Furthermore, their amplitude is only $1/2$, both for
the beam and the grating, since the Bloch vectors are confined to
the Northern hemisphere of the Bloch sphere, $0\leq
W_k^{(b,g)}\leq 1$. This explains the reduced amplitude of the oscillations in Fig. \ref{fig4}.

The analytic solution of this model given in the appendix yields,
for $N_b>N_g/2$, an oscillation period of
\begin{equation}
\tau_p=2\pi/\Omega_0=\frac{2}{U_0\sqrt{N_gN_b}}F((1/2+N_g/4N_b)^{1/2},\pi/2),
\label{period}
\end{equation}
where $F(k,\pi/2)$ is the complete elliptic integral of the first
kind \cite{schaum}.
\begin{figure}
\begin{center}
\includegraphics[width=8cm,height=11cm]{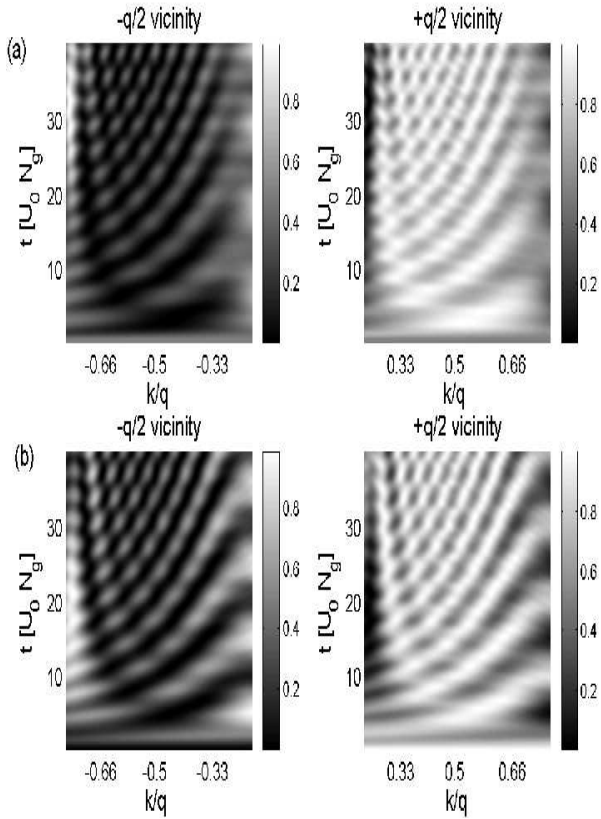}
\vspace{.3 cm}
 \caption{Atomic populations $n_k^{(b,g)}$ as a function of time (in units of
 $\tau_B=1/U_0 N_g$) for the case of short dephasing time 
 $\tau_{D}/\tau_{B}=1.0$,
 with $N_b/N_g=1.0$ and $\omega_{q}/N_{g}U_{0}=2$.
 The upper plots are for the grating and
 the lower plots for the beam.}
  \label{fig6}
\end{center}
\end{figure}

We have numerically calculated the Fourier transform of
$n_{k}^{(b,g)}(t)$ for $k_{F,b},k_{F,g}\ll q$ and
$\tau_D/\tau_B=2$ and confirmed that the populations of the
grating and the beam oscillate at the same frequency $\Omega_0$
independently of $k$. Fig. (\ref{fig5}) compares the period of
oscillations $\tau_p$ as a function of $N_b$ for fixed $N_g$ to
the analytical result of Eq. (\ref{period}), showing a very good
agreement. We find that the dominant frequency in the evolution of
$\rho_q(t)$ is approximately $\Omega_0/2$, in agreement
with the predictions of Bloch vector model. The small amplitude,
high-frequency oscillations in $\rho_q(t)$ are due to a residual
coupling to momentum states around $\pm 3q/2$.

The picture that we have developed indicates that the effects of
dephasing can only be suppressed when $\tau_D\gg \tau_B$. For
$\tau_B\gtrsim\tau_D$, the Bloch vectors of the grating no longer precess around
the $\hat{y}$-axis but rather around an axis at an angle
$\theta_k(t)=\arctan (U_0\sum_{k'}V_{k'}^{(b)}(t)/\delta\omega_k)$ relative to the
$\hat{z}$-axis at the ``generalized Rabi frequency''
$\sqrt{(U_0\sum_{k'}V_{k'}^{(b)})^2+\delta\omega_k^2}$. This again leads to the
spreading out of the Bloch vectors in the equatorial plane and
hence a collapse of the density grating. This is confirmed in Fig.
\ref{fig6}, which shows the dynamics of the populations forming
(a) the grating and (b) the beam for $N_b/N_g=1.0$ and
$\tau_D/\tau_B=1.0$. It illustrates clearly the frequency
dependence on $k$ familiar from the generalized Rabi oscillations
of two-level physics.

\begin{figure}
\begin{center}
\includegraphics[width=8cm,height=5cm]{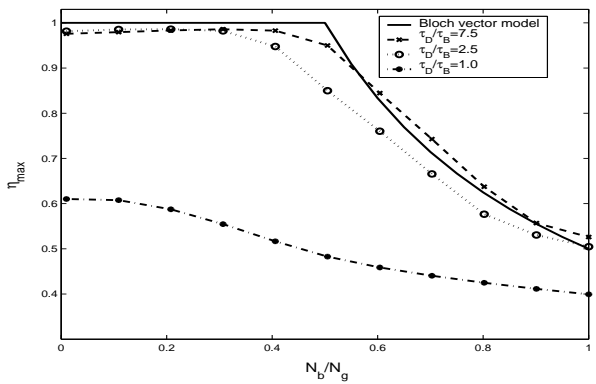}
\vspace{.3 cm}
 \caption{Diffraction efficiency $\eta_{max}$ as a function of $N_{b}/N_{g}$
 for $\tau_{D}/\tau_{B}=$ $7.5$ (dotted-line),
 $2.5$ (dashed line), and $1.0$ (dash-dot line),
 with $\omega_{q}/N_{g}U_{0}=2$. The solid line corresponds to
 maximum efficiency calculated from the Bloch vector model.
 }
  \label{fig7}
\end{center}
\end{figure}

The four-wave mixing efficiency can be quantified by the maximum
number of beam atoms that can be scattered by the grating.
Initially the beam atoms are scattered into negative momentum
states, hence the peak efficiency is given by the first maximum of
\begin{equation}
\eta(t)=\sum_{|k|<k_{F,b}} n^{(b)}_{k-q/2}(t)/N_{b}.
\end{equation}
Fig.~\ref{fig7} shows $\eta_{max}$ for different dephasing times.
We observe a rapid decrease in efficiency when $N_b/N_{g}>1/2$.
This result can be explained in terms of the Bloch vector model:
we show in the appendix that when $N_b>N_g/2$, the maximum number
of beam atoms that can be scattered into states around $-q/2$ is
$N_g/2$. In practice, the efficiency can exceed this maximum value
due to the residual coupling to states around $\pm 3q/2$, which
enhances the density modulation of the grating. Fig. \ref{fig7}
also illustrates the dramatic reduction in efficiency caused by
the dephasing of the grating. Even in that case, though, the
achievable efficiencies indicate that four-wave mixing should
still be observable when the Fermi gases used for the grating and
the beam contain comparable numbers of atoms.

\begin{figure}
\begin{center}
\includegraphics[width=8cm,height=6cm]{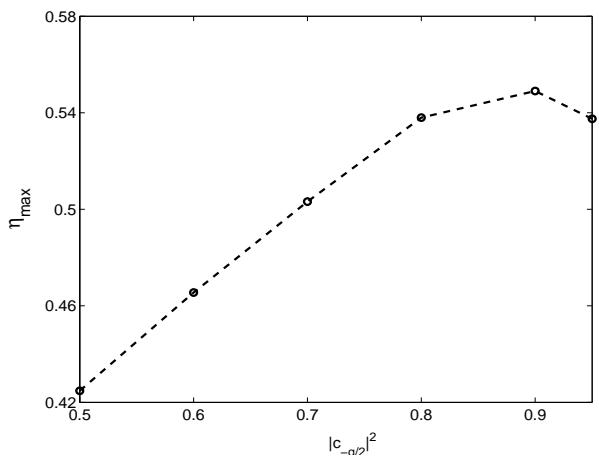}
\vspace{.3 cm}
 \caption{$\eta_{max}$ as a function of initial probability amplitudes
 for the grating $|c_{-q/2}|^2=1-|c_{q/2}|^2$,
 with $\tau_{D}/\tau_{B}=1.0$, $N_{b}/N_{g}=0.8$.}
  \label{fig8}
\end{center}
\end{figure}

Interestingly, one can improve the efficiency significantly by
choosing unequal probability amplitudes for the components of the
grating, that is, by replacing the initial state
$|\Psi^{(g)}\rangle$ of Eq. (\ref{initialconditions}) by the more
general superposition state
\begin{equation}
 \Ket{\Psi^{(g)}}=\prod_{|k|\leq
 k_{F,g}}\left( c_{-q/2} a^{\dag}_{k-q/2 \uparrow}+
 c_{q/2} a^{\dag}_{k+q/2 \uparrow}\right) \Ket{0}
\end{equation}
with $c_{-q/2}>c_{q/2}$ in our case. Fig.~\ref{fig8} plots the Bragg
efficiency for several values of $c_{q/2}$ and $c_{-q/2}$, showing
a maximum improvement of about $25\%$ as compared to the case of
equal initial amplitudes, $c_{-q/2} = c_{q/2}= 1/\sqrt{2}$. Even
though $c_{q/2}=c_{-q/2}=1/\sqrt{2}$ produces an initial density
grating with the largest possible amplitude, subsequent scattering
of beam atoms will reduce the size of the grating. In contrast,
while the amplitude of the density grating produced by an unequal
initial superposition state is less than optimal initially, the
subsequent scattering of beam atoms from $|k-q/2\rangle$ to
$|k+q/2\rangle$ increases $c_{q/2}$ while reducing $c_{-q/2}$,
resulting in a net increase over time in the grating amplitude.
\section{Conclusions}
In this paper we have investigated purely fermionic four-wave
mixing in a manner that accounts for the back-action of the
scattered beam on the density grating. Our numerical results
indicate that the lifetime for fermionic four-wave mixing can be
dramatically improved by making the number of atoms in the beam
comparable to the number of atoms that form the density grating.
Although the effects of the beam on the grating does reduce the
overall efficiency of the process, the efficiencies that we have
calculated indicate that it should be possible to convert in
excess of $10\%$ of the atoms in the incident beam into a fourth
scattered beam.

It is well known that four-wave mixing with bosons can lead to the
generation of squeezed states \cite{Yuen,Slusher}. Therefore a central
feature of future work will examine the statistical properties of
the scattered beam generated by the four-wave mixing to see if it
is possible to generate novel fermionic states. This will involve going beyond the
factorization ansatz in Eq. (\ref{factorize}) to calculate higher order moments for 
the particle hole operators. Phase conjugation should also be possible if one uses a
superfluid Fermi gas due to the presence of anomalous moments, 
$\langle a_{k \uparrow}a_{-k \downarrow}\rangle$, resulting from the formation of 
Cooper pairs.   

This work is supported in part by the US Office of Naval Research,
by the National Science Foundation, by the US Army Research
Office, by the National Aeronautics and Space Administration, and
by the Joint Services Optics Program. H. C. acknowledges the
support of the Studienstiftung des deutschen Volkes.
One of the author (T. M.) is supported by JSPS of Japan.

\section{Appendix}
In this appendix we solve the coupled Bloch equations for the beam
and grating with $\delta\omega_k = 0$. By introducing the
normalized vectors
\begin{eqnarray}
u^{(b,g)}=\sum_kU_k^{(b,g)}/N_{b,g}, \\
v^{(b,g)}=\sum_kV_k^{(b,g)}/N_{b,g},\\
w^{(b,g)}=\sum_kW_k^{(b,g)}/N_{b,g},
\end{eqnarray}
we obtain the equations of motion
\begin{eqnarray}
\dot{u}^{(b)}&=&-U_0N_g w^{(b)}v^{(g)}, \\
\dot{v}^{(b)}&=&U_0N_g w^{(b)}u^{(g)}, \\
\dot{w}^{(b)}&=&U_0N_g (u^{(b)}v^{(g)}-v^{(b)}u^{(g)}), \\
\dot{u}^{(g)}&=&-U_0N_b w^{(g)}v^{(b)}, \\
\dot{v}^{(g)}&=&U_0N_b w^{(g)}u^{(b)}, \\
\dot{w}^{(g)}&=&U_0N_b (u^{(g)}v^{(b)}-v^{(g)}u^{(b)}),
\end{eqnarray}
subject to the initial conditions $u^{(g)}(0)=w^{(b)}(0)=1$. Here,
the $\dot{}$ represents differentiation with respect to time. We
note that the constants of motion given by (\ref{constant1}) and
(\ref{constant2}) can be expressed as
\begin{equation}
N_bw^{(b)}+N_gw^{(g)}=N_b \label{pop}.
\end{equation}

For our particular initial conditions we have
$u^{(b)}(t)=v^{(g)}(t)=0$ for all times. This allows us to
parameterize the solution in terms of the angles $\theta_b(t)$ and
$\theta_g(t)$, of the beam and grating Bloch vectors relative to
the $\hat{z}$-axis in the $\hat{y}\hat{z}$- and
$\hat{x}\hat{z}$-planes, respectively,
\begin{eqnarray}
w^{(b)}(t)&=&\cos\theta_b(t), \\
v^{(b)}(t)&=&\sin\theta_b(t), \\
w^{(g)}(t)&=&\cos\theta_g(t), \\
u^{(g)}(t)&=&\sin\theta_g(t).
\end{eqnarray}
By introducing the sum and difference angles
\[
\theta_{\pm}=\theta_g \pm \theta_b -\pi,
\]
we obtain the uncoupled equations
\begin{equation}
\ddot{\theta}_{\pm}+\omega^2\sin\theta_{\pm}=0, \label{pendulum1}
\end{equation}
where $\omega=U_0\sqrt{N_gN_b}$. Eq. (\ref{pendulum1})
is the equation of motion for a plane pendulum \cite{marion}.

Since the initial conditions for these pendula are
$\theta_{\pm}(0)=-\pi/2$ and $\dot{\theta}_{\pm}(0)=\pm N_gU_0$, a
small amplitude linearization of the equations of motion is not
justified. However it is still possible to find a solution. We can
express the total "energy" for the pendula as
\[
\frac{1}{2}\dot{\theta}_{\pm}^2+2\omega^2\sin^2(\theta_{\pm}/2)=2\omega^2\kappa^
2,
\]
where
\[
\kappa^2=\frac{1}{2}+\frac{1}{4}\frac{N_g}{N_b},
\]
and $\theta_{\pm,0}=2\arcsin \kappa$ are the turning points for
the pendula. If $N_b<N_g/2$, then there are no turning points and
the pendula undergo complete revolutions about ``their'' axes. In
this case the Bloch vector for the beam undergoes complete
revolutions in the $\hat{y}\hat{z}$-plane as can be seen from Eq.
(\ref{pop}).

If, on the other hand, $N_b>N_g/2$, then we have the
oscillatory solution \cite{whittaker},
\begin{equation}
\sin(\theta_{\pm}(t)/2)=\pm\kappa \mbox{sn}(\omega(t\mp t_{0}),\kappa)
\end{equation}
where $\mbox{sn}$ is a jacobian elliptic function \cite{schaum}
and $t_{0}$ is determined by the initial condition,
$\kappa\mbox{sn}(\omega t_0,\kappa)=1/\sqrt{2}$. In this case the
beam Bloch vector has a turning point at $\cos\theta_b=1-N_g/N_b$,
so that the maximum number of beam atoms that can be scattered
into states around $-q/2$ is $N_g/2$. The period of the
oscillations for the elliptic function is given by Eq.
(\ref{period}).


\end{document}